# Automated and Secure Onboarding for System of Systems

SILIA MAKSUTI[1,2], ANI BICAKU[1,2], MARIO ZSILAK[1], IGOR IVKIC[1,3],
BÁLINT PÉCELI[4], GÁBOR SINGLER[4], KRISTÓF KOVÁCS[4], MARKUS TAUBER[1,5],
AND JERKER DELSING[2], (Member, IEEE)
[1]Cloud and Cyber-Physical Systems Security, University of Applied Sciences Burgenland, 7000 Eisenstadt, Austria
[2]Department of Computer Science, Electrical and Space Engineering, Luleå University of Technology, 971 87 Luleå, Sweden
[3]Department of Computing and Communications, Lancaster University, Lancaster LA1 4YW, U.K.
[4]evopro Innovation Ltd., 1116 Budapest, Hungary
[5]Research Studios Austria FG, 1090 Vienna, Austria

Corresponding author: Silia Maksuti (silia.maksuti@forschung-burgenland.at)

This work was supported in part by European Union (EU) Electronic Components and Systems for European Leadership (ECSEL) Joint Undertaking under Agreement 737459, Productive4.0 Project, and under Agreement 826452, Arrowhead Tools project, and in part by Investitionen in Wachstum und Beschäftigung–Europäischer Fonds für regionale Entwicklung (IWB-EFRE) (2014-2020), Measurement of IT-Security in Industry (MIT) 4.0 (FE02) Project.

**ABSTRACT** The Internet of Things (IoT) is rapidly changing the number of connected devices and the way they interact with each other. This increases the need for an automated and secure onboarding procedure for IoT devices, systems and services. Device manufacturers are entering the market with internet connected devices, ranging from small sensors to production devices, which are subject of security threats specific to IoT. The onboarding procedure is required to introduce a new device in a System of Systems (SoS) without compromising the already onboarded devices and the underlying infrastructure. Onboarding is the process of providing access to the network and registering the components for the first time in an IoT/SoS framework, thus creating a chain of trust from the hardware device to its hosted software systems and their provided services. The large number and diversity of device hardware, software systems and running services raises the challenge to establish a generic onboarding procedure. In this paper, we present an automated and secure onboarding procedure for SoS. We have implemented the onboarding procedure in the Eclipse Arrowhead framework. However, it can be easily adapted for other IoT/SoS frameworks that are based on Service-oriented Architecture (SoA) principles. The automated onboarding procedure ensures a secure and trusted communication between the new IoT devices and the Eclipse Arrowhead framework. We show its application in a smart charging use case and perform a security assessment.

## I. INTRODUCTION

Experts estimate that more than 64 billion devices will be part of the Internet of Things (IoT) by 2026 [1]. The diversity of these devices, their operating systems and running applications raises the challenge of providing generic security solutions either for the devices or the network where they are connected. Security of IoT devices, especially in industrial environments is of utmost importance since these devices interact with the physical world and have the potential to sabotage or even cause harm [2]. Addressing security is even more challenging when considering System of Systems (SoS) operating in different regions or platforms.

The associate editor coordinating the review of this manuscript and approving it for publication was Kashif Sharif.

SoS are large-scale integrated systems that are independently operable on their own but are networked together for a period of time to achieve a higher goal [3]. To take advantage of SoS, several industries are adopting existing technologies such as Service-oriented Architecture (SoA) to increase productivity, reduce operational costs and automatically carry out processes. To ensure that such systems are not compromised upon the arrival of new IoT devices, an onboarding procedure is needed.

In industrial environments device onboarding is usually done manually. The device ownership is transferred, configured on the network, and registered with the device owner in the IoT/SoS framework, which is costly and time consuming. The large number of devices and their complexity make the process of onboarding a challenging task, which can result in high security risks and vulnerabilities, e.g. device cloning, data tampering, etc. This process should be automated in order to reduce the effort to set up the device for onboarding and should fulfil security requirements defined by international security standards and best practice guidelines to mitigate these risks.

To address this issue, in this paper we present an automated onboarding procedure for IoT/SoS frameworks to design and build automation solutions. The aim of this procedure is to provide secure onboarding by establishing a chain of trust between a new hardware device, its hosted software systems and their provided services.

The onboarding procedure is implemented and evaluated using the Eclipse Arrowhead framework [4], which is based on SoA principles and facilitates the creation of local automation clouds. However, it can be easily adapted for other IoT/SoS frameworks that are based on SoA principles. The large number of devices, systems and services interacting with the Arrowhead local cloud raise security concerns, which require a specific solution for each device [5]. To ensure that the cloud is not compromised upon the arrival of new devices, the proposed onboarding procedure should be performed. The onboarding procedure introduces a process where devices, systems and services can be securely registered and automatically connected into the Arrowhead local cloud, allowing them to produce new services and consume already registered ones. One reason for this is to get reliable and trustworthy data, which can be used to make applications "smart", e.g. allowing them to respond to unsatisfactory situations [6]. The automated onboarding procedure ensures a secure and trusted communication between the new device and the Arrowhead local cloud.

In our previous work [7] we have provided an initial concept of the onboarding procedure. In this paper we provide an update of the existing systems involved in the onboarding procedure and present two additional systems, the Onboarding Contoller and the Certificate Authority (CA). The Onboarding Controller system is part of the local cloud chain of trust and is the first entry point to the local cloud. It accepts all the devices to connect with Hypertext Transfer Protocol Secure (HTTPS) protocol with the Onboarding service and on success provides: (i) the Arrowhead issued ‘‘onboarding'' certificate, and (ii) the endpoints of other services needed for the onboarding procedure. The CA system is responsible for signing any descendant certificates in an Arrowhead local cloud. All parties must trust the CA registered with the common name of its hosting local cloud. Thus, the onboarding procedure is extended to improve security and to address the complexity of devices and their needs.

We show the application of the automated and secure onboarding procedure in a smart charging use case. We perform a security assessment: (i) to identify potential threats, and (ii) assess the automated onboarding procedure against IEC 62443-3-3 standard to show how it can fulfil the security requirements of the standard for mitigating the identified threats. The results show that the proposed solution is compliant with the investigated security requirements and improves the performance by reducing the running time of the use case compared with manual onboarding.

The main contributions of this work are:

- Design and implementation of an automated onboarding procedure for establishing a chain of trust from the hardware device, to its hosted application systems and their provided services by creating a chain of certificates. A service cannot be registered without properly registering a system, a system cannot be registered without properly registering a device, and a device cannot be registered without a valid preloaded Arrowhead certificate, manufacturer certificate or shared key. The device, system and service unique identifiers and certificates are separately stored in the respective registries to increase security.
- Update of existing onboarding systems and introduction of two additional systems (Onboarding Controller, CA).
- Security assessment and time measurement of the proposed procedure in a representative industrial use case.

The remainder of this paper is structured as follows. Section II presents a number of surveys addressing IoT security and existing onboarding approaches and their limitations. Section III provides the necessary background for IoT security considerations. Section IV introduces the SoA basic principles. Section V introduces the automated and secure onboarding procedure, including a detailed description of involved systems. In section VI, we show the application of the onboarding procedure in a smart charging use case and perform a security assessment. Section VII outlines the findings and the future work.

## II. RELATED WORK

Several survey papers have been published addressing IoT domain. In [8] the authors evaluate a number of IoT frameworks against criteria such as architectural approach, industry support, standard-based protocols and interoperability, security, hardware requirements, governance and support for rapid application development. This evaluation can support academia and industry to identify the most suitable frameworks for their future projects. Another evaluation of IoT industrial frameworks is performed by authors in [9]. The IoT frameworks evaluated in this work include Arrowhead, Automotive Open System Architecture (AUTOSAR), BaSys, FIWARE, Industrial Data Space (IDS), Open Connectivity Foundation (OCF) and IoTivity. The evaluation highlights the general effort to solve problems such as security and interoperability. The authors emphasize the lack of industrial and automation requirements in some of the frameworks. In [10] the authors evaluate the security of IoT frameworks. They provide an introduction of some well-known IoT frameworks and conduct a comparative analysis of them based on security requirements, e.g. authentication, authorization and secure communication. The surveys presented in [11], [12], summarize the security threats and privacy concerns of IoT. In [12] the authors explore the most relevant limitations of IoT devices and their solutions, present the classification of IoT attacks, present mechanisms and architectures for authentication and access control, and analyze the security issues in different layers.

Many solutions already exist for IoT device onboarding. However, some of them are limited to the number of devices, type of devices or other dependencies (e.g. manufacturing or distribution). In this section we evaluate the most popular ones based on how they handle secure onboarding. Gupta and van Oorschot [13] have investigated onboarding and software update architectures for IoT devices. They highlight the need for an automatic procedure for secure software update. They explore the possibility of secure software update using public-key cryptography on a 8-bit micro-controller with 16MHz clock. As an example, a simple architecture with four components (IoT device, gateway device, smartphone application and software update provider) is considered. To build trust between all components, a secure onboarding protocol using key management is used. Also, they show the software update flow chart for the selected architecture. However, they consider only one possible scenario (shared key) and they do not investigate other scenarios such as, IoT devices with manufacturer certificates. In this work, we propose an onboarding procedure that shows this scenario and other possible scenarios in an industrial environment. In [14] the authors show the importance of device onboarding by proposing an onboarding approach for medical devices and healthcare services. They use the term 'tag' to extract information for the specific devices by providing three types of tags: (i) type I - device designer to provide hardware information, algorithm and data states, (ii) type II - signal analyzer to show the signal character and statistics, and (iii) type III - end user tags to check signal viability in the data steam, alerts and data quality. To illustrate the functionality, an onboard tagging use case considering a pulse oximeter is shown and tags type II are embedded in the data stream. However, they do not show the onboarding process of these tags in details or how other devices or medical services in a system of systems will benefit from this technology. Several studies, such as [15]–[17], etc., have been carried out for enabling efficient and secure onboarding for IoT devices. The authors in [15] noted that the device type is important when dealing with IoT devices in order to identify specific security vulnerabilities. Knowing the type and the vulnerabilities can help decide if the device will join or not the network. This approach can mitigate or reduce the impact of attacks, but in an industrial environment is not always possible. The secure onboarding procedure presented in this work will categorize the device type based on the credentials they use to interact with the local cloud and this will determine if they are allowed to join the network. Kumar *et al.* [16] propose a software based solution that can be used to securely onboard devices with microphone. This procedure is done via a voice command to onboard multiple IoT devices. Other automatic solutions such as Intel zero touch device onboarding [17] already exist, but these are solutions that depends on specific capabilities of the device. In [16] the device should have a microphone and in [17] the device should have an embedded microchip named EPID (Enhanced Privacy ID) that needs to be installed during manufacturing.

The automated onboarding procedure proposed in this paper supports devices with three credential types: (i) device with Arrowhead issued certificate, (ii) device with manufacturer certificate, and (iii) device with shared key. However, the procedure itself is not device dependent, any device having an Arrowhead issued certificate, a manufacturer certificate, or a shared key can be securely onboarded.

### III. BACKGROUND
The rapid growth of connected devices, ranging from small sensors to production devices, increases the risk of security threats. In this section we introduce a number of security requirements, e.g. identity management, authentication and authorization, secure communication protocols and secure elements, which should be considered in IoT/SoS frameworks. Additionally, we provide an insight how existing frameworks address these security requirements.

#### A. SYSTEM OF SYSTEMS
As defined by ISO/IEC/IEEE 21839, the System of Systems (SoS) are *" a set of systems or system elements that interact to provide a unique capability that none of the constituent systems can accomplish on its own. Note: Systems elements can be necessary to facilitate the interaction of the constituent systems in the system of systems"* [18]. They are distributed systems composed of several components and have several characteristics that distinguish them from traditional systems, such as: (i) operational independence, (ii) managerial independence, (iii) evolutionary development, (iv) emergent behaviour, (v) geographic distribution and other characteristics including autonomy, belonging, connectivity, diversity and emergence [19]. These characteristics should be taken into consideration when designing an SoS. Since these systems evolve during the time, security is a major challenge since a SoS can integrate systems with different security requirements where systems with a low security level can compromise systems requiring a high level of security.

#### B. INTERNET OF THINGS
As defined by ISO/IEC JTC 1, the Internet of Things (IoT) is *"an infrastructure of interconnected objects, people, systems and information resources together with intelligent services to allow them to process information of the physical and the virtual world and react"* [20]. Thus, IoT is a platform used to connect heterogeneous and distributed things embedded with electronics, software and sensors to the internet enabling them to collect and exchange vast amounts of data. These data are then analyzed to build business intelligence and new business models to improve user experience. Currently, there is no standardised architecture for IoT that is agreed universally, because different users have different requirements. However, a basic architecture includes the perception layer, where sensors are used to sense and gather information from the environment, the network layer, which is responsible for connecting things and for transmitting and processing sensor data, and the application layer, which is responsible for delivering application specific service to the users. Some well-known IoT frameworks are Eclipse Arrowhead, Amazon Web Services (AWS) IoT, Azure IoT Suite, Kura, etc. IoT allows the creation of System of Systems (SoS), which are independently operable on their own, but are networked together for a period of time to achieve a higher goal, e.g. costs, performance, robustness. Since these systems evolve during time, security is a major challenge because in a SoS can be integrated systems with different security requirements.

### C. IDENTITY MANAGEMENT

In ISO/IEC 24760-1, identity management is defined as *"processes and policies involved in managing the life cycle and value, type and optional metadata of attributes in identities known in a particular domain"* [21]. In IoT, identity management should be able to identify devices, systems and services and their access to confidential data, rather than identifying only people. Thus, it is of utmost importance to establish a naming system for IoT devices, systems and services, as well as to create a process for registering these entities in a secure way. Additionally, the focus in SoA changes from traditional systems to reusable services, thus identity management is a cross-cutting concern. In [22] the authors present a blueprint for a service-oriented identity management architecture featuring interoperability by applying existing standards. An authentication service is used to issue security tokens, which enables the web services for single sign-on, and an authorization service is used for separation of concerns. In [23] the authors propose an approach for identity and access management in the context of SoA by defining a domain-specific language (DSL) for role-based access control (RBAC) that allows for the definition of identity and access management policies for SoA. In [24] the authors propose a model to manage the integration of identity, authentication and authorization modules based on formal policy-based methods. In [25] is presented a new naming convention for the Eclipse Arrowhead framework following the requisites and characteristics defined in the system of systems integration. The naming convention represents a renovated vision of the identification of devices, systems and services. Thus, provides a trusted chain of connections, describing who is hosting what information and enabling security policy implementation. Identity management in IoT is performed by exchanging identifying information between the entities for first time connection. This process is susceptible to eavesdropping, which can lead to man-in-the-middle attack, and thus can jeopardize the whole IoT framework. Hence, security techniques such as authentication and authorization should be integrated in this process.

### D. AUTHENTICATION AND AUTHORIZATION

Authentication ensures that a person, device, system or service is the one claimed. Authentication factors include something you know, e.g. password, PIN code, something you have, e.g. tokens, certificates, debit cards, and something you are, e.g. bio-metrics. The certificate based authentication works via asymmetric cryptography. Each party needs to have a valid certificate, which is signed by a trusted and known parent certificate. The trusted certificate confirms the identify of the presented certificate, known also as the chain of trust concept. Figure 1 shows a chain of trust with three certificates. The certificate with the highest authority is known as root certificate. The root certificate signs its descending certificates, confirming their identity. Any certificate between the root certificate and the end-entity certificate is called an intermediate certificate.

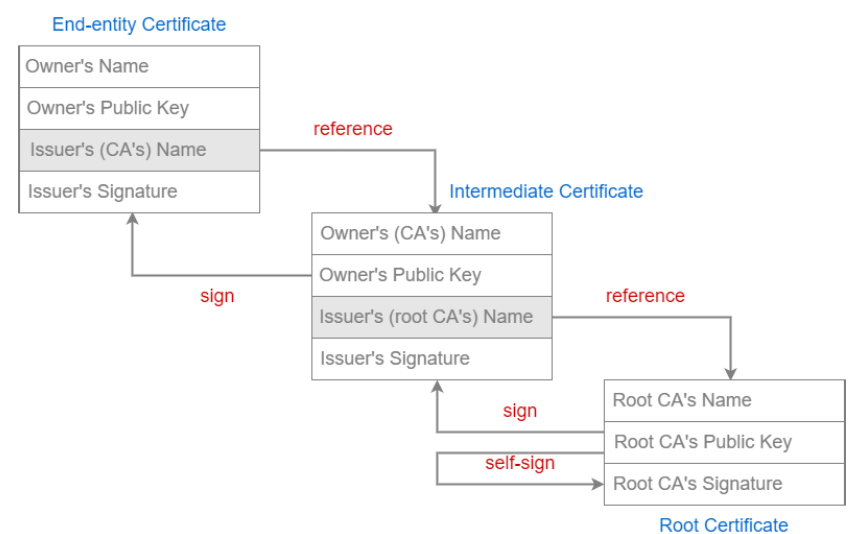

**FIGURE 1.** Chain of trust including root certificate, intermediate certificate and end-entity certificate.

In Arrowhead framework the Certificate Authority system, which will be introduced later on in the paper, is used to sign any descendent certificate in an Arrowhead local cloud. Amazon Web Services (AWS) IoT [26], a cloud platform for IoT released by Amazon, provides three ways of verifying identity, X.509 certificates, AWS Identity and Access Management (IAM) users, groups, and roles, AWS Cognito identities. Thus, each device, connected to the AWS IoT, is authenticated using one of these methods chosen by the end-user. Azure IoT Suite [27], a platform released by Microsoft, is composed of a set of services that enable end-users to interact with their IoT devices. In order to establish a secure connection between devices and Azure IoT Hub, the handshake process is encrypted using Transport Layer Security (TLS). An identity proof in terms of X.509 certificate is sent to the targeted device, which has a unique identity key at deployment time. The device then authenticates itself to Azure IoT Hub by sending a token, which contains a HMAC-SHA256 signature string that is a combination of the generated key along with a user-selected device identity. Kura, an Eclipse IoT project, provides a framework for IoT gateways that run M2M applications [28]. Kura uses secure sockets and Eclipse Paho clients to handles the majority of data communication via MQTT protocol.

According to NIST SP 800-82, authorization is *"the right or a permission that is granted to a system entity to access a system resource"* [29]. Thus, it is the process of granting access privileges and is determined by applying policy rules, e.g. access control mechanism, to the authenticated person, device, system or service. In Arrowhead framework the Authorization system, which will be introduced later on in the paper, is used to provide authentication, authorisation and optionally accounting of service interactions. In AWS IoT, the authorization process is policy-based. Thus, it can be applied by either mapping rules and policies to each certificate or by applying IAM policies. Azure IoT uses Azure Active Directory (AAD) [30] to provide a policy-based authorization model for data stored in the cloud. Kura has a service component, which manages security policies.

### E. SECURITY IN ECLIPSE ARROWHEAD

In this section we briefly explain how identity management, authentication and authorization is handled in the Eclipse Arrowhead framework and the corresponding systems. Paniagua *et al.* [25] propose a naming convention for the Arrowhead framework, which is used as a guideline for entity identification within the local cloud. The naming convention provides a trusted chain of connections, describing who is hosting what information and enabling security policy implementation. The naming convention covers services to local clouds. The Arrowhead local cloud can contain many devices. Each device can contain one or more systems. Each systems can contain one or more services. Thus, identifiers will be used to connect the hosted device/system/service with the hosting local-cloud/device/system. Identifiers used in the Arrowhead naming convention follow the same structure as identifiers in DNS-SD and RFC-6335 recommendations [31].

The local cloud name is important in a SoS scenario. The identifier of a local cloud should be connected to its Gatekeeper system, which is the system in charge of the connection with other local clouds [32].

The local cloud identifier is as follows:

_gatekeepersystemname._InterCloudNegotiations._protocol._transport._InterCloudNegotiations:port

The device identifier is as follows:

_devicename._localcloudname._interface._macprotocol._macaddress

The system identifier is as follows:

_systemname._devicename._protocol._transport._domain

The service identifier is as follows:

_servicename._sysname._protocol._transport_domain:port

In order to connect a new device to the Arrowhead local cloud, the device has to be authenticated. The technique used for authentication in Arrowhead is X.509 certificates [33]. They are digital certificates, which should be issued by a trusted party, a certificate authority. These certificates are SSL/TLS-based to ensure secure authentication. In this paper we introduce a new Arrowhead core system, the Certificate Authority (CA), which is part of the certificate hierarchy in the Arrowhead framework shown in Figure 2. The CA system is the highest authority in the local cloud and is responsible for signing any descending certificates, e.g. onboarding certificates, device certificates, and system certificates, which are needed for the onboarding procedure. All services in the local cloud must trust the CA of the local cloud. The CA itself may be signed by a central Arrowhead consortium, establishing a chain of trust and allowing different Arrowhead local clouds to interconnect with each other.

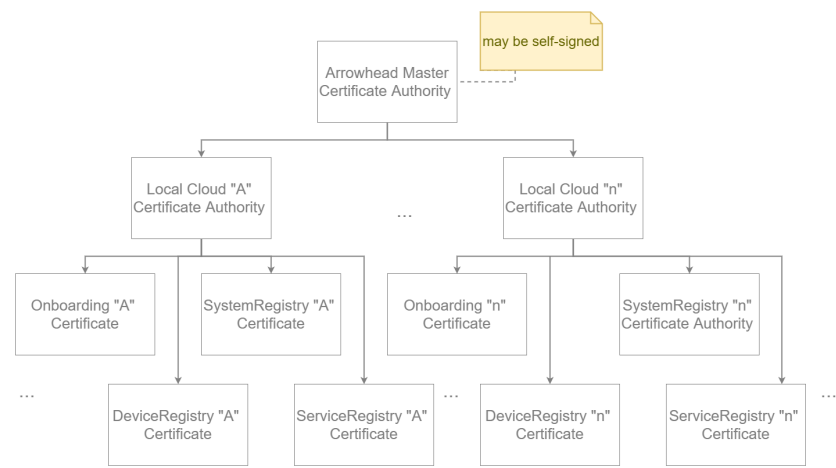

**FIGURE 2.** Certificate hierarchy in Arrowhead. The Certificate Authority system is the highest authority in the local cloud and is responsible for signing any descending certificates needed for the automated onboarding procedure.

The authorization process is policy-based. The AuthorizationControl service, produced by the Authorization system, is responsible for enforcing all authorization related policies. A policy may define that no systems with a specific operating system may be allowed in the cloud. Another policy may define that a service (e.g. temperature collector service) may only query the ServiceRegistry about existing temperature services but not e.g. power regulating services.

### F. SECURE COMMUNICATION PROTOCOLS

In industrial IoT environments, there is a need to connect legacy devices and sensors. Thus, not only IP- and Ethernet-based protocols, but also serial-based protocols must be connected. This adds integration complexity and security concerns. Some of the most used IoT application-layer protocols include Hypertext Transfer Protocol (HTTP), Constrained Application Protocol (CoAP) and Open Platform Communications United Architecture (OPC-UA). HTTP is used for distributed, collaborative, hypermedia information systems [34]. It is defined as a request-response protocol in the client-server computing model. HTTP Secure (HTTPs) is an extension of the HTTP, which is used for secure communication on the Internet. In HTTPS, the communication protocol is encrypted using TLS. CoAP is a web transfer protocol based on REpresentational State Transfer (REST) [35]. CoAP is often used as a lightweight alternative to HTTP for use with constrained devices. CoAP does not provide security in itself, but it uses Datagram Transport Layer Security (DTLS) at the transport layer, in order to secure all CoAP messages. DTLS provides data confidentiality and integrity, authentication, non-repudiation and anti-replay protection for CoAP communication. CoAP with DTLS support is known as secure CoAP (CoAPs) [36]. OPC-UA is the new standard of the OPC

Foundation providing interoperability in process automation and beyond. By defining abstract services, OPC-UA provides SoA for industrial applications – from factory floor devices to enterprise applications [37]. Since OPC-UA runs over Transmission Control Protocol (TCP), with optional HTTPs encoding, it is not suitable for low-power devices. Instead, User Datagram Protocol (UDP) uses fewer resources and provides low-power operation.

Messaging protocols such as Message Queue Telemetry Transport (MQTT), Advanced Message Queuing Protocol (AMQP), and Extensible Messaging and Presence Protocol (XMPP) are also used. To ensure communication security MQTT, AMQP and XMPP, are used over TLS. Some application-layer protocols, such as CoAP and MQTT, are more appropriate for running on constrained devices. Others, like XMPP, are recommended for the communication between gateways and servers over the Internet [38].

However, industrial IoT environments continue to use more industry-specific protocols. IEC 61850-9-2 Sampled Values (SV) [39], IEC 61850-8-1 Generic Object Oriented Substation Event (GOOSE) [40], Manufacturing Message Specification (MMS) [41], Modbus [42], etc., provide the core communication mechanisms for industrial environments such as power utilities, manufacturing, and transportation.

### G. HARDWARE ROOT OF TRUST

Above mentioned software-based security mechanisms are not sufficient to protect against security threats, since data may be collected by potentially untrusted devices. Thus, it is important to add an additional hardware-based security layer. Each device should have a hardware ''secure element''. The secure elements provide tamper resistant storage for holding and protecting the key from any kind of attack, even including physical access to the device. One example is Trusted Platform Module (TPM) [43], which is a hardware chip designed to enable commodity computers to achieve greater levels of security and is specified by the Trusted Computing Group (TCG) industry consortium. TPMs are manufactured by chip producers, including Atmel, Broadcom, Infineon, STMicroelectronics, Winbond, etc. TPMs provide integrity protection and a root of trust for the devices. Furthermore, TPMs provide a standardized interface, which makes it very easy to integrate them in any device.

## IV. SERVICE-ORIENTED ARCHITECTURE

The continuous growth of IoT and its benefits in multiple industries will require emerging tools and technologies to meet organizations and consumers needs. To take advantage of IoT, several industries are adopting existing technologies such as Service-oriented Architecture (SoA) to increase productivity, reduce operating costs and automatically carry out processes. The automated and secure onboarding procedure is implemented and evaluated using the Eclipse Arrowhead framework, which is based on SoA principles. Thus, in this section we provide an overview of this technology.

SoA is a technology that allows applications to be registered as services. Thus, it is about information exchange between a service producer and a service consumer as shown in Figure 3. SoA provides automation of industrial systems based on the following principles: (a) loose coupling, which supports autonomy and distributed services, (b) late binding, which makes possible to use the information any time by connecting to the correct resources, and (c) lookup, which can be used to discover already registered services.

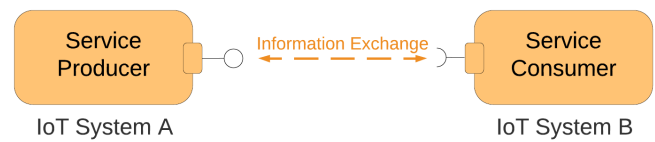

**FIGURE 3.** Service-oriented Architecture. SoA is about information exchange between a service producer and a service consumer.

### a: LOOSE COUPLING
Two SoA systems do not need to know about each other at design time to allow a run time data exchange. The identification of available services is established at run time making use of a service registry system and its discovery mechanisms as shown in Figure 4. A new SoA service will register itself in the service registry and it will be discoverable by any other service in the network.

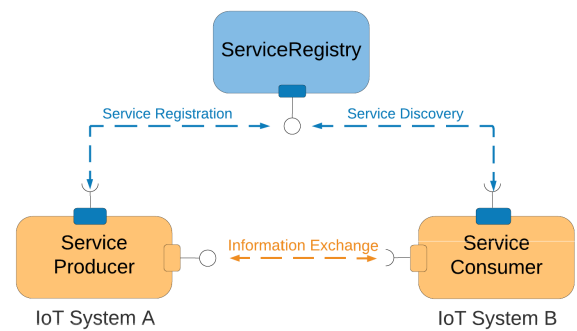

**FIGURE 4.** SoA loose coupling supports autonomy and distributed services.

### b: LATE BINDING
The exchange of data between two systems is established in run time as shown in Figure 5. The run time coupling is initiated by an orchestration mechanism, which provides the endpoint of the selected producer to the requesting consumer. If necessary, the authorization mechanisms is consulted to check if the service consumer system can be authenticated and authorized to consume the requested service.

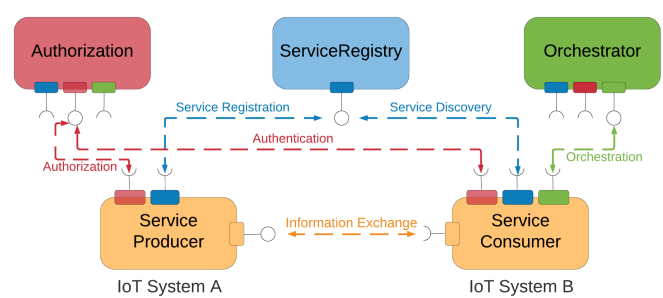

**FIGURE 5.** SoA late binding makes possible to use the information any time by connecting to the correct resources.

### c: LOOKUP

In a SoA environment the data exchange can be initiated by a service consumer requesting data, which is known as pull behaviour. A pull behaviour can be controlled by a timer at the service consumer by creating data pulling of a sensor every 100 ms. The data exchange can also be initiated by a producer that knows about conditional data request, which is known as pull behaviour. This is initiated by a data subscription under certain criteria. For e.g. a pressure sensor will push its pressure reading service to a consumer whenever the pressure reading is higher than 2 bar, data is then pushed from the producer to the consumer.

However, the principles of SoA have not been designed to primarily address security. Originally, SoA implementation was associated with SOAP (Simple Object Access Protocol). More recently, developers prefer to use lightweight REST services instead. REST uses HTTP to obtain data and perform operations, and supports SSL authentication to achieve secure communication. REST is a stateless protocol, e.g. each HTTP request contains all necessary information, which means that neither the service producer nor the service consumer are required to retain any data to satisfy the request.

Despite the benefits, REST application programming interfaces (APIs) are prone to known API attacks such as Man In The Middle (MITM) attacks, API injections (e.g. cross site scripting (XSS), SQL injection (SQLi)), Distributed Denial of Service (DDoS) attacks. To protect against such threats, security best practices, such as authentication and authorization, should be in place. It is important to know who is using the services of a SoA-based environment in order to control access. This can be done using a variety of standards, some established such as X.509 certificates, and some new such as WS-Security [44]. To ensure that IoT devices, systems and services are authenticated and authorized to connect to a SoA-based IoT/SoS framework, we propose a secure onboarding procedure, which is described in the next section.

## V. AUTOMATED ONBOARDING PROCEDURE

The automated and secure onboarding procedure is needed when a new device produced by any vendor (e.g. Siemens, Infineon, Bosch, etc.) wants to interact with an IoT framework, as shown in Figure 6.

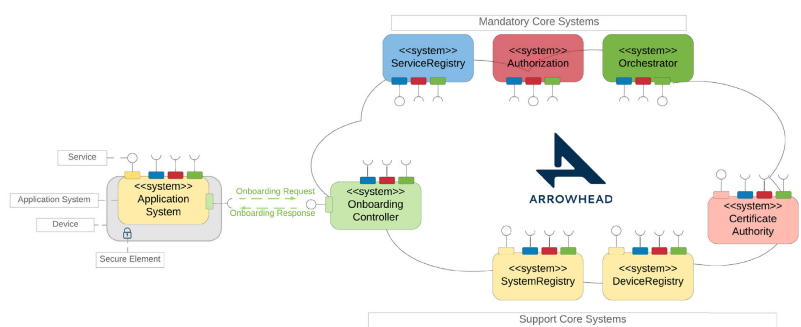

**FIGURE 6.** Automated onboarding procedure ensures that devices, systems and services are properly authenticated and authorized to connect to the Arrowhead local cloud. In the first step, the new device communicates via https with the Onboarding Controller system.

To assure that the framework is not compromised upon the arrival of this new device, it is important to establish a chain of trust from the new hardware device, containing a secure element (e.g. TPM), to its hosted application systems and their services. Thus, the onboarding procedure makes possible that IoT devices, systems and services are authenticated and authorized to connect to an IoT/SoS framework. Low-level IoT devices have low processing power and small memory sufficient for dedicated tasks. It is challenging to deploy public-key cryptography and to deliver software updates through the internet for these constrained devices [45].

The onboarding procedure is validated using the Eclipse Arrowhead framework,[1] which facilitates the creation of local automation clouds used to enable local real time performance and security, interoperability, simple and cheap engineering and scalability through multi cloud interaction. The source code of the onboarding systems and their documentation can be found in the open-source Arrowhead Github repository.[2] The results should easily be adapted for other IoT frameworks, which are based on SoA principles.

### A. ONBOARDING PROCEDURE SYSTEMS

In the following we provide an overview of the systems involved in the onboarding procedure.

### 1) ONBOARDING CONTROLLER SYSTEM

The Onboarding Controller system has to be part of every local cloud where trust at device, system and service level is required. Thus, it is a core system and belongs to the Arrowhead local cloud chain of trust. The Onboarding Controller system: (i) is the first entry point to the local cloud, e.g. accepts all devices to connect via the Onboarding service, (ii) has a certificate for the *https* communication with the device, and (iii) (optionally) the certificate is provided by a public CA (e.g. Verisign). On success, the system provides: (i) an Arrowhead issued ''onboarding'' certificate, and (ii) the endpoints of the DeviceRegistry, SystemRegistry, ServiceRegistry and Orchestrator systems.

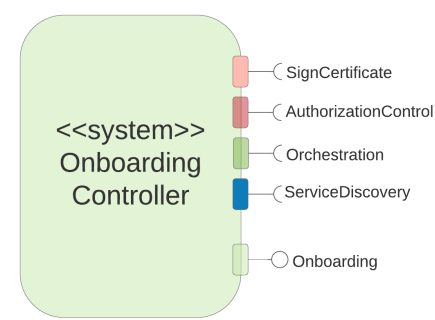

**FIGURE 7.** Onboarding Controller is a support core system in Arrowhead framework. It produces the Onboarding service, which grants onboarding with an Arrowhead issued certificate, manufacturer certificate or a shared secret.

The Onboarding Controller system shown in Figure 7 consumes the ServiceDiscovery, Orchestration, AuthorizationControl and SignCertificate services and provides the Onboarding service. Its functionalities are shown in Table 1.

[1]https://www.arrowhead.eu/
[2]https://github.com/eclipse-arrowhead/core-java-spring

**TABLE 1.** Onboarding Functions.

| Function | URL Path | Method | Input | Output |
|---|---|---|---|---|
| certificate | "/certificate/name" | POST | Onboarding With Name | Onboarding With Name Resposne |
| certificate | "/certificate/csr" | POST | Onboarding With CSR | Onboarding With CSR Response |
| sharedSecret | "/sharedsecret/name" | POST | Onboarding With Name | Onboarding With Name Response |
| sharedSecret | "/sharedsecret/csr" | POST | Onboarding With CSR | Onboarding With CSR Response |

The *OnboardingWithCsr* input contains a Base64 encoded Certificate Signing Request as required by the CA system of the local cloud. The *OnboardingWithName* input contains a common name that will be in the certificate. Authentication of the client will be done through a shared secret (e.g. a password without user id as defined in RFC2617 [46] Basic Authentication) or through mutual authentication of a trusted manufacturer certificate (see RFC5246 [47]). The *OnboardingResponse* contains a number of fields e.g. the indication if the operation is successful, the URIs of the endpoints of DeviceRegistry, SystemRegistry, ServiceRegistry and Orchestrator systems, Base64 encoded signed certificates returned by CA system, the algorithm of the key e.g. RSA, the format of the key e.g. X.509, etc.

Figure 8 shows the use cases that represent the actors and their interaction with the Onboarding Controller system.

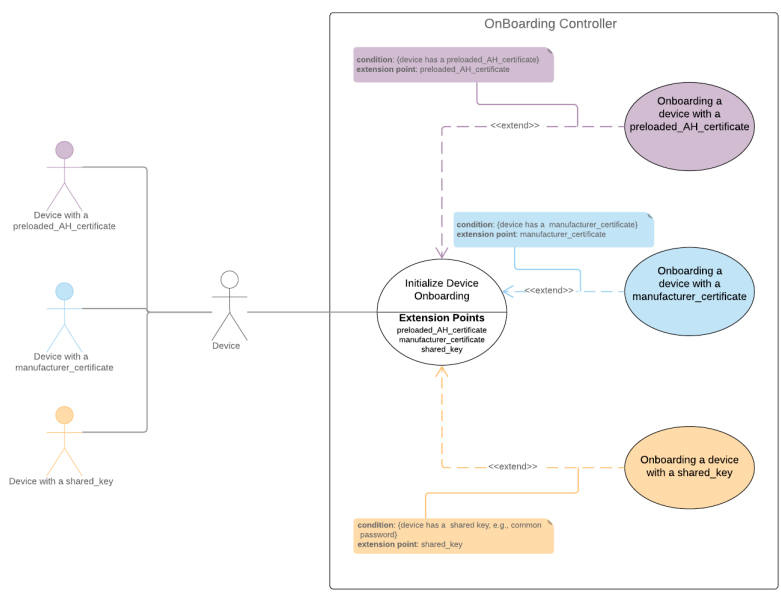

**FIGURE 8.** Onboarding Controller supports three use cases: (i) onboarding with preloaded Arrowhead certificate, (ii) onboarding with manufacturer certificate, and (iii) onboarding with a shared secret.

The actors can be devices with different credentials: (i) preloaded Arrowhead certificate, (ii) manufacturer certificate, and (iii) shared secret.

2) DeviceRegistry SYSTEM

The DeviceRegistry system is used to provide a local cloud storage holding the information on which devices are registered within a local cloud, meta-data of these registered devices, including a list of the systems that are deployed in each of them. The DeviceRegistry system holds for the Arrowhead local cloud unique device identities. The DeviceRegistry system shall be accessible using different SoA protocols (e.g. REST, CoAP, MQTT). As shown in Figure 9, the DeviceRegistry system consumes the three mandatory core services of Arrowhead, the SignCertificate service provided by CA and provides the DeviceDiscovery service.

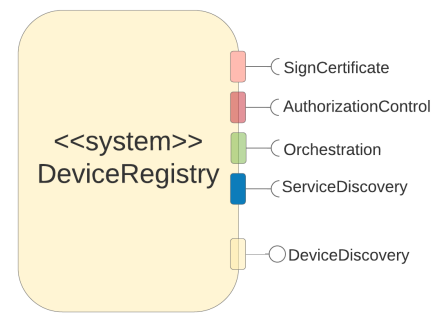

**FIGURE 9.** DeviceRegistry is a support core system in Arrowhead framework. It consumes the Arrowhead mandatory core services, the SignCertificate service, and produces the DeviceDiscovery service.

The DeviceDiscovery service provides the functionalities shown in Table 2.

**TABLE 2.** DeviceDiscovery Functions.

| Function | URL Path | Method | Input | Output |
|---|---|---|---|---|
| Register | "/register" | POST | Device Registry Entry | Device Registry Entry |
| Unregister | "/unregister" | DELETE | Device Name, MAC Address | OK |
| Query | "/query" | POST | Device Query Form | Device Query List |
| Onboard | "/onboarding/name" | POST | Onboarding With Name | Onboarding With Name Response |
| Onboard | "/onboarding/csr" | POST | Onboarding With Csr | Onboarding With Csr Response |

The *register* function is used to register a device, which contains a symbolic name as well as a physical endpoint. The instance parameter represents the endpoint information that should be registered. The *unregister* function is used to unregister a device that no longer should be used. The instance parameter contains information necessary to find the device to be removed. The *query* function is used to find and translate a symbolic device name into a physical endpoint, IP address and a port. The query parameter is used to request a subset of all the registered devices in the DeviceRegistry system based on a specified criteria. The *onboard* function is an extension of the *register* function and is used during the onboarding of a device. It is the only function which accepts the ''onboarding'' certificate. Homogeneously to the Onboarding Controller system, the onboard function consumes the SignCertificate and returns an Arrowhead device certificate. The *DeviceRegistryEntry* contains a number of fields e.g. the Arrowhead device object that is provided, the name and the mac address of the Arrowhead device, endofValidity e.g. an ISO 8601 format date-time, and metadata. Metadata should be provided using key pairs such as, encode=syntax, e.g. encode=xml, compress=algorithm, e.g. compress=exi, semantics=XX, e.g. semantics=senml.

Figure 10 shows the use cases that represent the actors and their interaction with DeviceRegistry system.

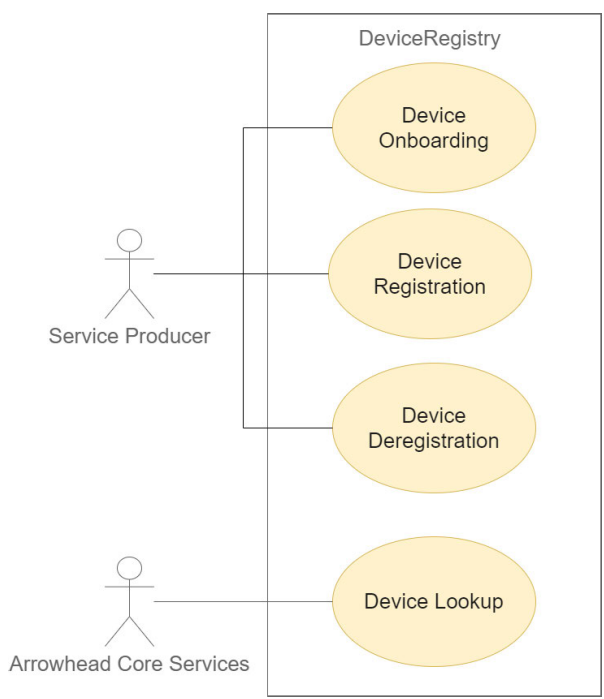

**FIGURE 10.** DeviceRegistry supports four use cases: (i) device onboarding, (ii) device registration, (iii) device deregistration, and (iv) device lookup.

### 3) SystemRegistry SYSTEM

The SystemRegistry system is used to provide a local cloud storage holding the information on which systems are registered within a local cloud, meta-data of these registered systems and the services these systems are designed to consume. The SystemRegistry holds for the Arrowhead local cloud unique system identities for systems deployed within it. As shown in Figure 11, the SystemRegistry system consumes the three mandatory core services of Arrowhead, the SignCertificate service produced by CA, and produces the SystemDiscovery service.

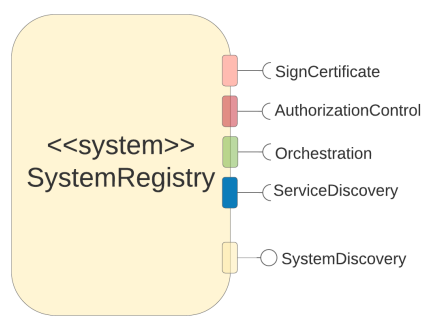

**FIGURE 11.** SystemRegistry is a support core system in Arrowhead framework. It consumes the Arrowhead mandatory core services, the SignCertificate service, and produces the SystemsDiscovery service.

The SystemDiscovery service provides the functionalities shown in Table 3.

The *register* function is used to register a system, which contains a symbolic name as well as a physical endpoint. The instance parameter represents the endpoint information that should be registered. The *unregister* function is used to unregister a system that no longer should be used. The instance parameter contains information necessary to find the system to be removed. The *query* function is used to find and translate a symbolic system name into a physical endpoint, IP address and a port. The query parameter is used to request a subset of all the registered systems in the SystemRegistry system based on a specified criteria. The *onboard* function is an extension of the *register* function and is used during the onboarding of a system. It is the only function which accepts the device certificate. Homogeneously to the Onboarding Controller and DeviceRegistry systems, the onboard function consumes the SignCertificate and returns an Arrowhead system certificate.

**TABLE 3.** SystemDiscovery Functions.

| Function | URL Path | Method | Input | Output |
|---|---|---|---|---|
| Register | "/register" | POST | System Registry Entry | System Registry Entry |
| Unregister | "/unregister" | DELETE | System Name, Address, Port | OK |
| Query | "/query" | POST | System Query Form | System Query List |
| Onboard | "/onboarding/name" | POST | Onboarding With Name | Onboarding With Name Response |
| Onboard | "/onboarding/csr" | POST | Onboarding With Csr | Onboarding With Csr Response |

The *SystemRegistryEntry* contains a number of fields e.g. the Arrowhead system object that is provided, the name and IP address of the Arrowhead system, the port where the provided system can be consumed, authentication information e.g. if the communication is secure provides the public key of the systems' certificate, the Arrowhead device that is providing the system, endOfValidity, metadata, etc.

Figure 12 shows the use cases that represent the actors and their interaction with SystemRegistry system.

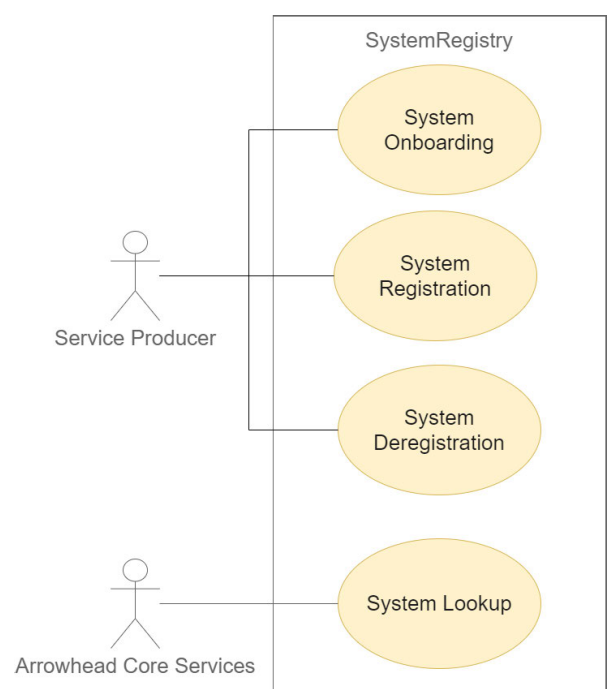

**FIGURE 12.** SystemRegistry supports four use cases: (i) system onboarding, (ii) system registration, (ii) system deregistration, and (iv) system lookup.

#### 4) ServiceRegistry SYSTEM

The ServiceRegistry system provides the database, which stores information related to the currently actively offered services within the local cloud. In this generation of the framework, this system is implemented twice, using two different database technologies: (i) DNS-SD BIND server with a Java-bases REST interface, and (ii) MySQL database using the same Java-based REST interface. The purpose of this system is therefore to allow application systems to register the services they offer at the moment, making this announcement available to other application systems on the network, to remove or update their entries when it is necessary, and to utilize the lookup functionality of the registry to find public core system service offerings in the network, otherwise the Orchestrator has to be used. As shown in Figure 13, the ServiceRegistry system produces the ServiceDiscovery service.

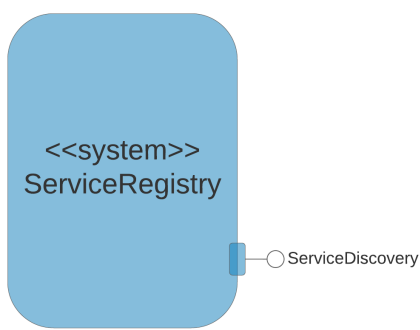

**FIGURE 13.** ServiceRegistry is a mandatory core system in Arrowhead framework. It produces the Service Discovery service.

The ServiceDiscovery service provides the functionalities shown in Table 4.

**TABLE 4.** ServiceDiscovery Functions.

| Function | URL Path | Method | Input | Output |
|---|---|---|---|---|
| Register | "/register" | POST | Service Registry Entry | Service Registry Entry |
| Unregister | "/unregister" | DELETE | Address, Port, Service Definition, System Name | OK |
| Query | "/query" | POST | Service Query Form | Service Query List |

The *register* function is used to register services. The services will contain various metadata as well as a physical endpoint. The various parameters are representing the endpoint information that should be registered. The *unregister* function is used to unregister service instances that were previously registered in the ServiceRegistry. The instance parameter is representing the endpoint information that should be removed. The *query* function is used to find and translate a symbolic service name into a physical endpoint, for example an IP address and a port. The query parameter is used to request a subset of all the registered services fulfilling the demand of the user of the service. The list contains service endpoints that fulfill the query.

In the latest generation of the framework, the lookup functionality of services is integrated within the orchestration process. Therefore, when an application system is looking for a service to consume, it shall ask the Orchestrator system via the Orchestration service to locate one or more suitable service providers and help establish the connection based on metadata submitted in the request. Direct lookup from application systems within the network is not advised in this framework generation, due to security reasons. However, the lookup of other application systems and services directly is not within the primary use, since access will not be given without the Authorization JSON Web Token. The use of the Token Generation is restricted to the Orchestrator for general system accountability reasons.

Figure 14 shows the use cases that represent the actors and their interaction with ServiceRegistry system.

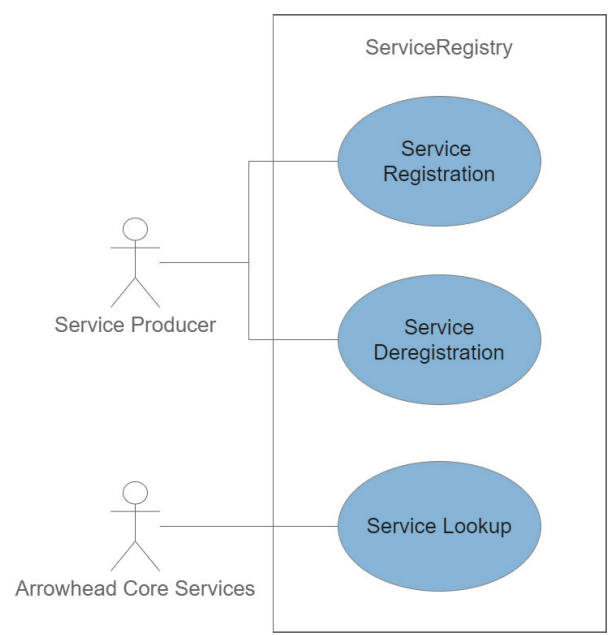

**FIGURE 14.** ServiceRegistry supports three use cases: (i) service registration, (ii) service deregistration, and (iii) service lookup.

The above mentioned registries, use a basic three-tier architecture: (i) The presentation tier named *Controller.java, which transforms the view into domain specific objects and vice versa. Each RegistryController contains three functions, *query*, which searches for an entity with e.g. a specific name (common name as shown in its certificate), *register*, which stores an entity in the database, and *unregister*, which removes an entity from the database. (ii) The application tier is named *Service.java, which is responsible for any business logic and extensive validation. (iii) The data tier is taken from the Arrowhead common project and is named *Repository.java. The Repository classes are generic interfaces using Spring Boot Data for implementation, They allow to deal with any entity in the Arrowhead code. Arrowhead uses OpenAPI (formerly known as swagger [48]) to enrich the documentation of its REST methods.

#### 5) AUTHORIZATION SYSTEM
The Authorization system has a database that describes which application system can consume what services from which application systems, intra-cloud access rules, and a database that describes which other local clouds are allowed to consume what services from this cloud, inter-cloud authorization rules. The Authorization system provides the AuthorizationControl service and the Token Generation service for allowing session control within the local cloud, as shown in Figure 15. The purpose of the Token Generation functionality is to create session control functionality through the core systems. The output is JSON Web Token that validates the service consumer system, when it wants to access the service from another application system. This token shall be primarily generated during the orchestration process and only released to the service consumer when all affected core systems are notified and agreed to the to-be-established service connection. The AuthorizationControl service provides two different interfaces to look up authorization rights: (i) intra-cloud authorization, which defines an authorization right between a consumer and provider system in the same local cloud for a specific service, and (ii) inter-cloud authorization, which defines an authorization right for an external cloud to consume a specific service from the local cloud.

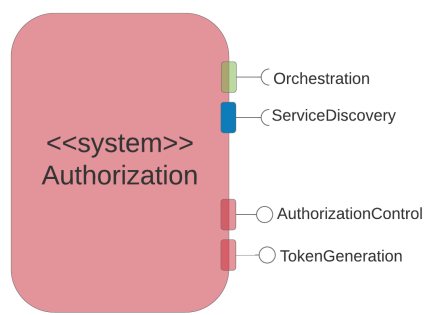

**FIGURE 15.** Authorization is a mandatory core system in Arrowhead framework. It produces two services, AuthorizationControl and Token Generation.

The AuthorizationControl service provides one functionality as shown in Table 5. *Get Public Key* returns the public key of the Authorization core service as a Base64 encoded text. This service is necessary for service providers if they want to utilize the token based security.

**TABLE 5.** AuthorizationControl Functions.

| Function | URL Path | Method | Input | Output |
|---|---|---|---|---|
| Get Public Key | "/publickey" | GET | - | Public Key |

The Authorization system, in line with all core systems, utilizes the X.509 certificate Common Name naming convention in order to work.

#### 6) ORCHESTRATOR SYSTEM

The Orchestrator system provides run-time binding between application systems. The purpose of the Orchestrator system is to provide application systems with orchestration information, where they need to connect to. The outcome of the Orchestration service includes rules that will tell the application system what service provider system(s) it should connect to and how. Such orchestration rules include accessibility information details of a service provider (e.g. network address and port), details of the service instance within the provider system (e.g. base URL, Interface Design Description specification and other metadata), authorization-related information (e.g. access token and signature) and additional information that is necessary for establishing connection. As shown in Figure 16, the Orchestrator system provides two core services, the Orchestration service and the OrchestrationStoreManagement service, and should consume at least the two mandatory core services.

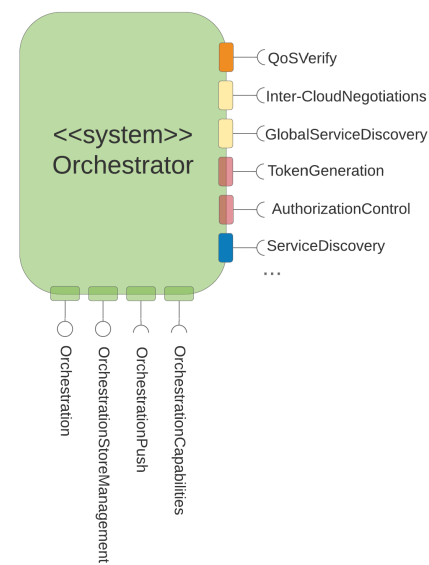

**FIGURE 16.** Orchestrator is a mandatory core system in Arrowhead framework. It produces OrchestrationStrore and OrchestrationStroreManagement services.

The Orchestrator system can be used in two ways: (i) using predefined rules coming from the Orchestrator Store database to find the appropriate providers for the consumer, and (ii] dynamic orchestration, in which the core service searches the whole local cloud to find matching providers.

#### 7) CERTIFICATE AUTHORITY (CA) SYSTEM

The CA is responsible for signing any descendant certificates in an Arrowhead local cloud. All parties must trust the CA registered with the common name of its hosting local cloud. The certificate of the CA may be signed by a central authority (e.g. Arrowhead Consortium), so, the chain of trust can be established allowing different local clouds to interconnect with each other in a secure manner. The Certificate Authority system consumes the mandatory core services and provides the SignCertificate service, as illustrated in Fig. 17.

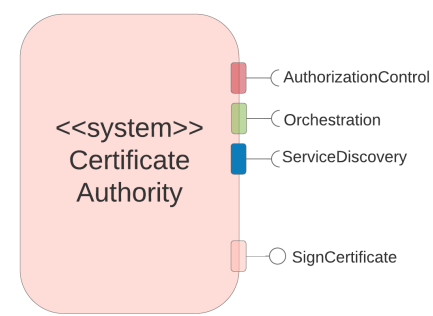

**FIGURE 17.** CertificateAuthority is a support core system in Arrowhead framework. It consumes the Arrowhead mandatory core services and produces the SignCertificate service.

The SignCertificate service issues signed certificates for requester entities inside a local cloud. The requester entity has to construct a *Certificate Signing Request* (CSR) in compliance with [49] and send it to the CA. The CA verifies the signature inside the CSR. If the signature verification is successful, then the CA generates and sends back a signed certificate for the requester entity. Using this certificate, the requester entity is able to communicate in secure manner with the systems inside the local cloud. The SignCertificate service consists of a single function as shown in Table 6.

**TABLE 6.** SignCertificate Functions.

| Function | URL Path | Method | Input | Output |
|---|---|---|---|---|
| SignCertificate | "/getSigned Certificate" | POST | Certificate SigningRequest | Certificate SigningResponse |

### B. ONBOARDING PROCEDURE SEQUENCE DIAGRAM

As shown in Figure 18, the use cases in which an external actor (e.g. new device) interacts with the Arrowhead local cloud during the onboarding procedure include: (i) initialize device onboarding via the Onboarding service, (ii) onboard and register a device in the DeviceRegistry system via the DeviceDiscovery service, (iii) onboard and register a system in the SystemRegistry system via the SystemDiscovery service, (iv) (optionally) register a service in the ServiceRegistry system via the ServiceDiscovery service, and (v) start normal operation (e.g. service lookup, service consumption).

The sequence diagram in Figure 19 shows the interaction of a new device with the Arrowhead local cloud during the onboarding procedure. This a valid procedure once the device has been connected to the private network of the local cloud. Thus, the new device must be connected to the Onboarding Controller system from outside the local cloud or from a demilitarized zone (DMZ) within the local cloud. The device hosting the Onboarding Controller needs to have two network interfaces, one connected to the private network and one connected to the DMZ or the open internet. Once allowed to be part of the private network, the device a) can physically be connected to the private network by an operator, or b) if in DMZ, a reconfiguration of the firewall can be made by the Onboarding Controller system. As mentioned above, the device can have different credentials e.g. a preloaded Arrowhead certificate, a manufacturer certificate or a shared secret. However, in this paper we show only the sequence diagram of a device with a manufacturer certificate stored on a TPM, which provides the highest security as its private key cannot be read from outside. Following is provided a step-by-step description of the onboarding procedure:

1) We are assuming that the new device has a *manufacturer certificate*, which is stored in a secure element, e.g. TPM. The device sends an onboarding request to the Onboarding Controller system, which contains the manufacturer certificate. The communication with the Onboarding service is done using https protocol.

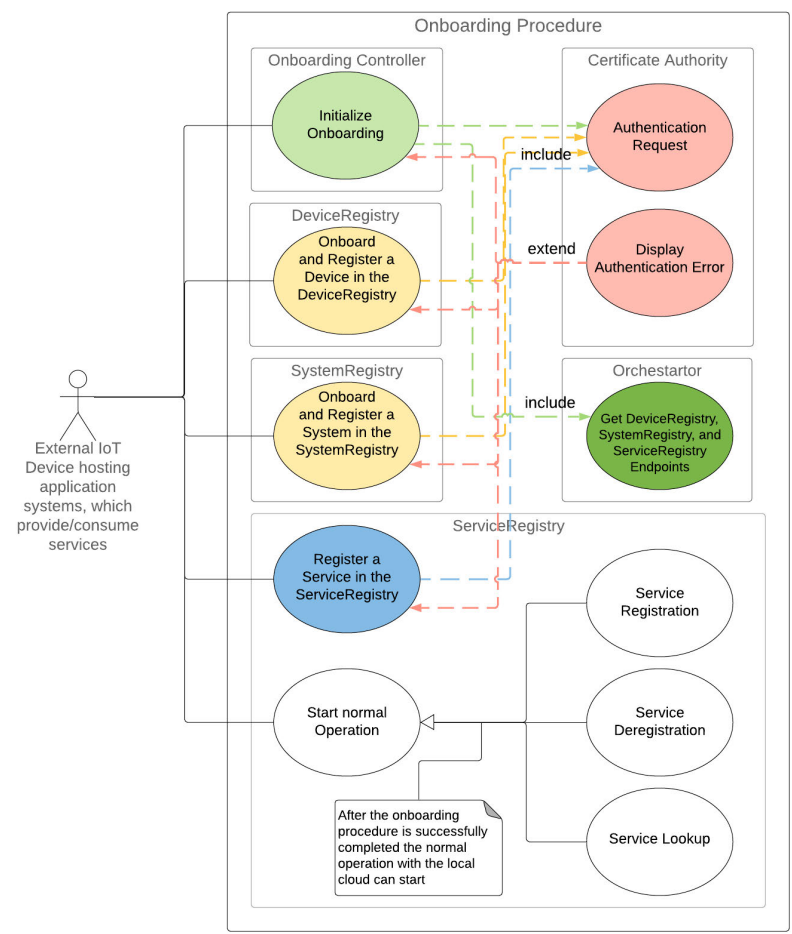

**FIGURE 18.** Onboarding use cases. During the automated onboarding procedure the new device interacts with a number of Arrowhead systems for initializing device onboarding and for registering device, systems and services.

2) The Onboarding Controller system verifies the manufacturer certificate with the CA system, which, on success, sends the Arrowhead "onboarding" certificate.
3) The Onboarding Controller system gets the endpoints of DeviceRegistry, SystemRegistry, and ServiceRegistry from Orchestrator system.
4) The Onboarding Controller system returns the Arrowhead "onboarding" certificate and the endpoints of DeviceRegistry, SystemRegistry, ServiceRegistry and Orchestrator to the device.
5) Using the obtained endpoints the new device begins the registration with the DeviceRegistry system through the DeviceDiscovery "onboard" interface authenticating itself with the *Arrowhead onboarding certificate*.
6) The DeviceRegistry either: (i) stores intermediate certificates issued by CA and based on them authenticates the device without using any CA service, or (ii) verifies the received Arrowhead onboarding certificate with the local cloud CA.
7) On success, the Arrowhead local cloud CA issues the Arrowhead "device" certificate.
8) The DeviceRegistry registers the device and returns the Arrowhead "device" certificate to the device.
9) The system on the new device begins the registration with the SystemRegistry system through the SystemDiscovery "onboard" interface authenticating itself with the *Arrowhead device certificate*. This procedure is repeated for each system hosted on the device.

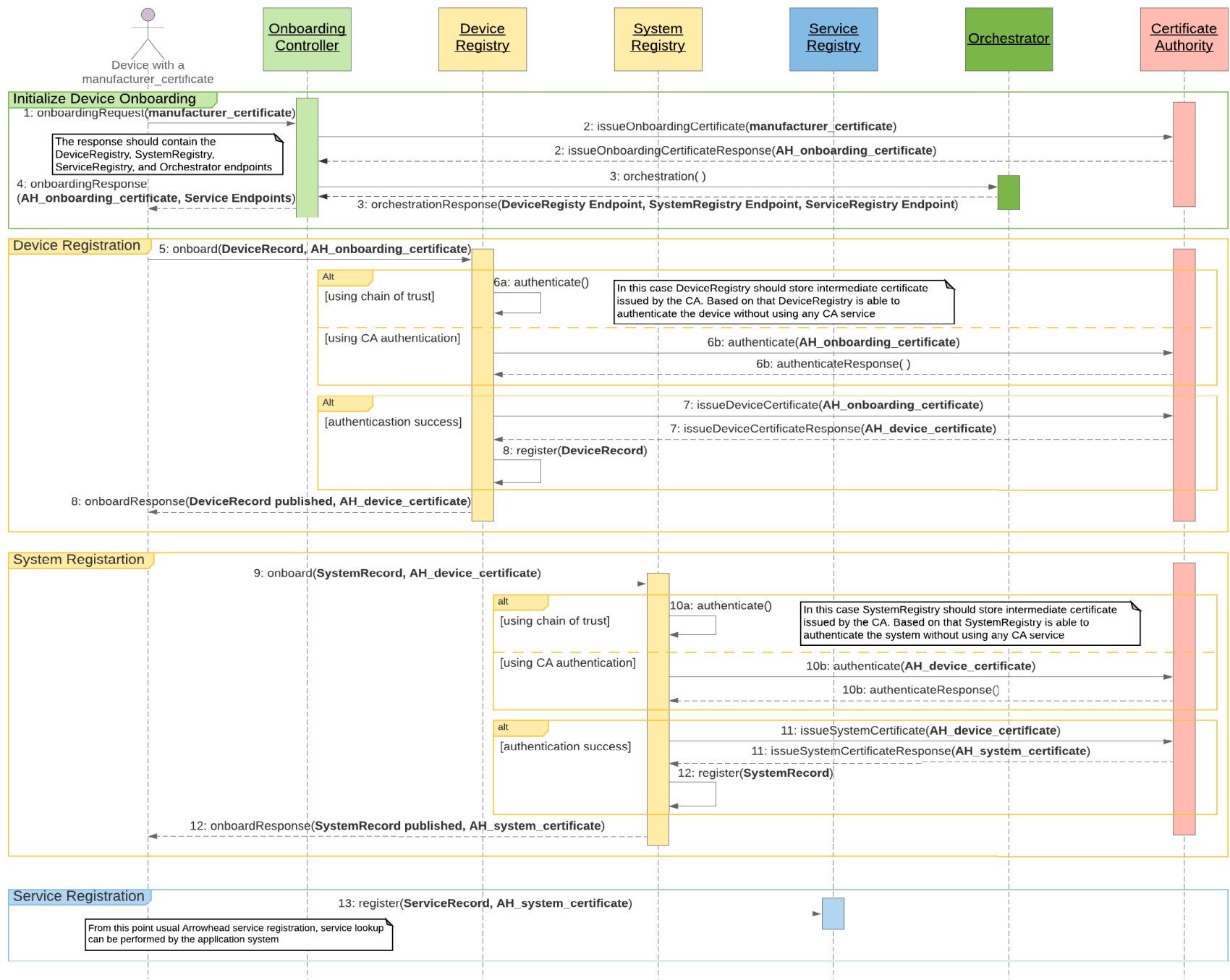

**FIGURE 19.** The sequence diagram of the automated onboarding procedure for a new device with a manufacturer certificate.

10) The SystemRegistry either: (i) stores intermediate certificates issued by CA and based on them authenticates the system without using any CA service, or (ii) verifies the received Arrowhead ''device'' certificate with the local cloud CA.
11) On success, the local cloud CA issues Arrowhead ''system'' certificates for each registered system.
12) The SystemRegistry registers the system and returns the Arrowhead ''system'' certificate.
13) The system begins the registration of its produced service in the ServiceRegistry using the ''register'' interface of the ServiceDiscovery, authenticating it with the *Arrowhead system certificate*. Since a system owns its own data the acceptance of the system certificates will transfer the right to manage consumption of produced services to the Authorization system. This procedure is repeated for each service produced by the system.

From this point normal operation can start, e.g. service lookup can be performed by the application system. In the latest generation of the framework, the lookup functionality of services is integrated within the orchestration process. This onboarding procedure provides a chain of trust from the new hardware device, to its hosted software systems and their services by using a chain of certificates, manufacturer certificate, Arrowhead onboarding certificate, Arrowhead device certificate and Arrowhead system certificate.

## VI. SMART CHARGING USE CASE

The Arrowhead local cloud can be considered as a specific corporate sub-network, to which increased security and QoS requirements apply. As such, connecting any new devices to the local cloud means a use case for the onboarding procedure. Trivial use cases include the deployment of intelligent actuators to production lines or new sensor nodes to data acquisition networks. Here, the automated onboarding procedure shall eliminate the need for any manual installation and configuration efforts.

A different scenario is the administration and authentication of mobile devices, which may enter and leave local clouds on an ad-hoc basis. An example for that is the interaction of a charging station and an electrical vehicle (e-vehicle) arriving to charge. The use case setup shown in Figure 20 is composed of a charging station, an e-vehicle and an Arrowhead local cloud.

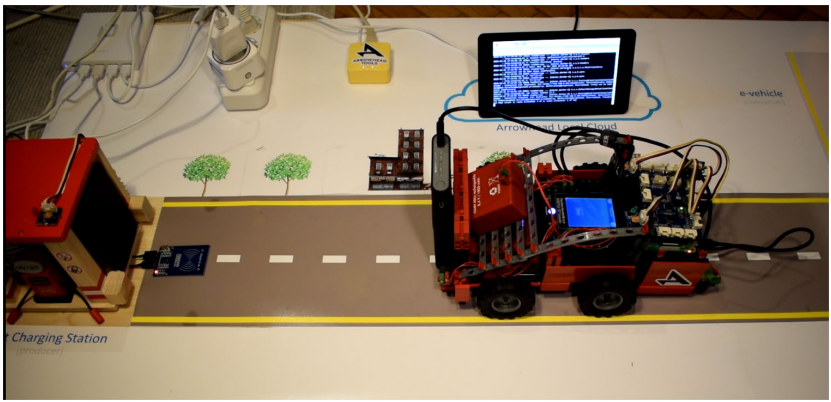

**FIGURE 20.** Smart charging use case setup. Secure onboarding of charging station and e-vehicle in the Arrowhead local cloud.

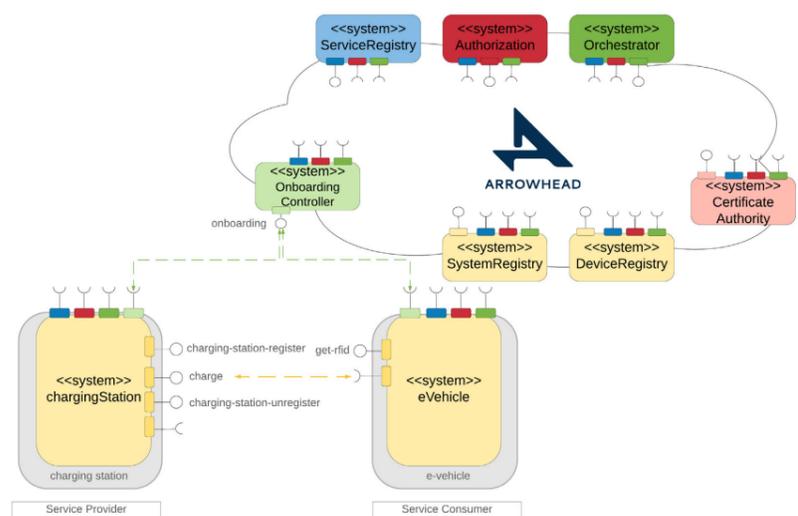

**FIGURE 21.** Smart charging use case devices, application systems and services. Secure onboarding of charging station and e-vehicle in the Arrowhead local cloud.

The Arrowhead framework defines three maturity levels (ML) for application systems: ML3, the application system implements the consumption/production of services without external components, ML2, the application system implements the consumption/production of services by using a software adapter, and ML1, the application system uses dedicated hardware with software responsible for wrapping the application system with Arrowhead framework compliant services. In our use case, dedicated hardware is used to implement the consumption/production of services.

The **charging station (service provider)** is composed of: (i) an inductive charger used to refuel the battery and simulate the charging of the e-vehicle, (ii) a voltcraft measuring device to control when the charger is supplied with power, (iii) a RFID reader to identify the service consumer, and (iv) a Raspberry Pi that runs Arrowhead, controls the voltcraft and the RFID reader. The **e-vehicle (service consumer)** is composed of: (i) a Fischertechnik TXT controller to control the engine and sensors of the e-vehicle, (ii) a battery, which is used to power the Raspberry Pi and will be charged by the charging station, (iii) a RFID chip card to identify the service consumer to the service provider, and (iv) a Raspberry Pi that runs Arrowhead. The **Arrowhead local cloud**, including mandatory core systems and onboarding systems, runs in a Raspberry Pi. An info-screen is used to display information regarding Arrowhead and status of the use case and a WiFi router is used for the communication.

Figure 21 extends Figure 6 by showing the devices, application systems and services specific to the use case.

The charging Station system running on charging station device produces three services: *charging-station-register* service, which allows other systems to register in the station, *charging-station-unregister* service, which allows other systems to unregister from the station, and *charge* service, which allows other systems to request charging from the station. The *charge* service contains the consumers RFID as parameter. The eVehicle system running on e-vehicle device produces the *get-rfid* service, which is consumed by the charging station to verify the RFID.

Both, charging station and e-vehicle shall be securely onboarded on the Arrowhead local cloud to communicate with each other. Once the e-vehicle gets into the WiFi range, it sends an onboardingRequest to the *onboarding* service of the Onboarding Controller system. If succeeded, the e-vehicle goes through the automated onboarding procedure steps described in Section V-B. It receives an onboarding certificate to register the device (e-vehicle), a device certificate to register its hosted system (eVehicle) and a system certificate to register the corresponding service (*get-rfid*).

After onboarding is completed, the e-vehicle registers itself with the charging station by consuming the *charging-station-register* service. Then it drives to the charging station. Once the e-vehicle is in close proximity, the RFID reader of the charging station reads the RFID of the e-vehicle. The charging station triggers its own *charge* service with the e-vehicles' RFID and charges it on success. The charge has a fixed duration. The registered application system reports battery data from the CAN bus of the e-vehicle, which is used to individualize the charging session. The e-vehicle disengages once the charge is done and unregisters itself from the charging station.

In our use case a RFID reader and chip is used to trigger the charging of the e-vehicle. However, in real-world scenarios such a trigger can be realized through distance sensors, weight sensors or even cameras. The onboarding procedure enables fully-automated user authentication at the charging station making any type of manual identification (RFID, mobile app) unnecessary. A short video of this use case can be found in YouTube (*www.youtube.com/watch?v=F-mG9s2ttT8&ab_channel=EclipseArrowhead*).

### A. TIME MEASUREMENT

The Eclipse Arrowhead framework relies on certificates and mutual authentication as one of its security concepts. Thus, every device, system and service must have an individual certificate issued by the local cloud. Creating and deploying those certificates can be seen as mandatory setup costs. The onboarding procedure reduces such costs through its built-in automated certificate creation.

To measure the time it takes to perform the automated onboarding procedure, the smart charging use case is divided into three main parts: (i) onboarding, which consists of initializing device onboarding, persisting the certificate to the flash drive, and device and system registration, (ii) charging, which consists of starting/initializing the service, preparing the use case and executing the use case, and (iii) deregistration, which deregisters the service, system, and device from the Arrowhead local cloud. We have executed the smart charging use case several times to extract the average duration of each part. The onboarding part took on average 14.9 seconds (35.91% of total time). The charging part, which contains a fixed 10 seconds charge in the execution phase, took on average 23.2 seconds (55.84% of total time). The deregistration part took on average 3.4 seconds (8.25% of total time). Thus, the total time needed for running the use case when using automated onboarding procedure is on average 41.5 seconds.

These results are compared with the time needed for running the use case when using manual onboarding. We have measured the time of manual certificate creation based on the official development procedure available in Arrowhead Github repository[3] and the deployment to their respective devices. In order to compare the execution of the automated onboarding procedure with the manual one, we mapped the manual certificate creation to the initialize device onboarding part and the certificate deployment to the persisting the certificate part. The certificate creation and deployment time for the manual onboarding took on average 205 seconds. Device and system registration of the onboarding part, as well as the charging and deregistration parts, took the same time as with the automated onboarding procedure. The total time needed for running the use case when using manual onboarding procedure is on average 233 seconds.

Thus, the automated onboarding procedure improves the performance since it reduces the time needed to run the smart charging use case. The results can be seen in Figure 22.

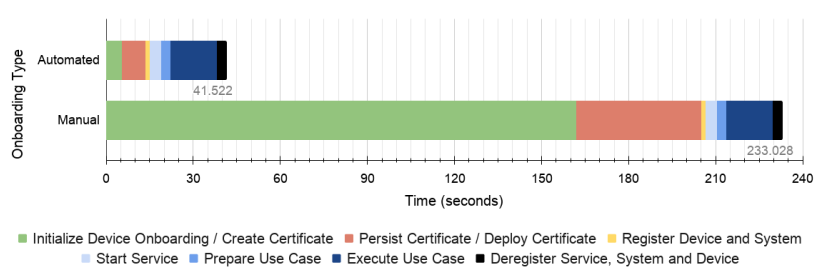

**FIGURE 22.** Time Measurement. Time needed for running the smart charging use case when using automated and manual onboarding procedure. Standard deviation is negligible, thus not included in the graph.

Generated certificates, independent thereof if they are manually created or through the automated onboarding procedure are valid for a long period of time. The certificate creation guide recommends a validity of 10 years, while the automated onboarding procedure, in its default configuration, produces a validity of 1 year. It is expected that the onboarding time is negligible compared to repeated executions of the use case throughout the validity of the certificates.

[3]https://github.com/eclipse-arrowhead/core-java-spring/blob/master/documentation/certificates/create_client_certificate.pdf

### B. SECURITY ASSESSMENT

In this section, we present the security assessment of the smart charging use case. The security assessment is performed to determine how effectively the automated onboarding procedure meets specific security requirements.

To identify and mitigate potential security threats, we have used Microsoft STRIDE as a threat modeling tool. STRIDE categorizes different types of threats into Spoofing (S), Tampering (T), Repudiation (R), Information Disclosure (I), Denial of Service (D), and Elevation of Privilege (E) [50]. The process of threat modeling according to STRIDE can be divided into a number of steps: (i) creating an application diagram to identify the assets and their interactions, (ii) identifying threats, and (iii) mitigating threats. Thus, as a first step we have created an application data flow diagram for the use case, shown in Figure 23.

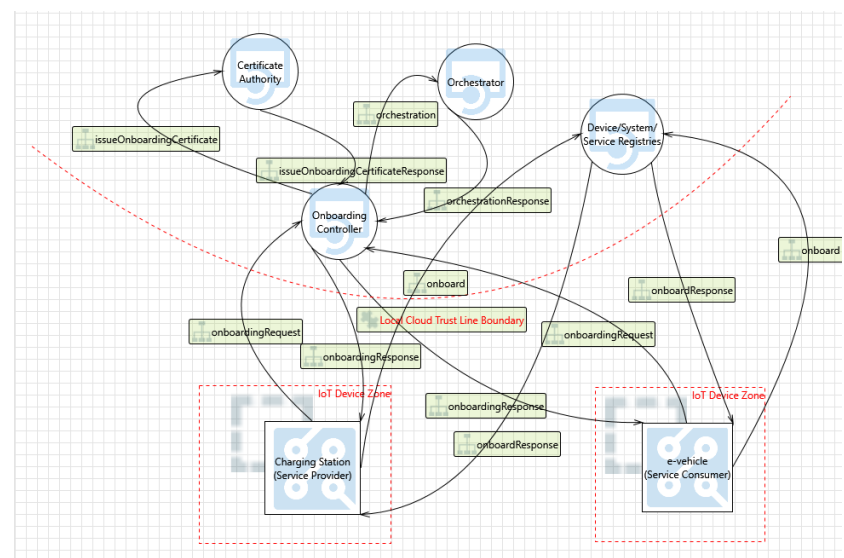

**FIGURE 23.** Data Flow Diagram (DFD) of the smart charging use case.

Based on the data flow diagram, 88 threats are identified. In Table 7, for each threat category is shown the affected security objective and the number of the identified threats.

**TABLE 7.** Threat modeling results using Microsoft STRIDE.

| Security Objectives | STRIDE Threat Category | Nr. of Threats |
|---|---|---|
| Authentication | S - Spoofing | 8 |
| Integrity | T - Tampering | 32 |
| Non-Repudiation | R - Repudiation | 8 |
| Confidentiality | I - Information Disclosure | 24 |
| Availability | D - Denial of Service | 0 |
| Authorization | E - Elevation of Privilege | 16 |

According to the results, the most affected security objectives in this use case are integrity and confidentiality since the highest number of threats is detected in Tampering (T) and Information Disclosure (I) categories. No threats are detected in Denial of Service (D) category, one reason for this could be that the services are not exposed in the Internet.

In terms of the Eclipse Arrowhead framework, all involved systems exchange information by means of services. Thus, each interaction between the systems involved in the smart charging use case provides the same threats. Since it would be redundant to show how the proposed onboarding procedure mitigates all 88 identified threats, where the only difference would be the system producer and consumer, we show one representative micro use case and its identified threats. The micro use case, interaction between the charging station (service provider) and the Onboarding Controller system, is shown in Figure 24.

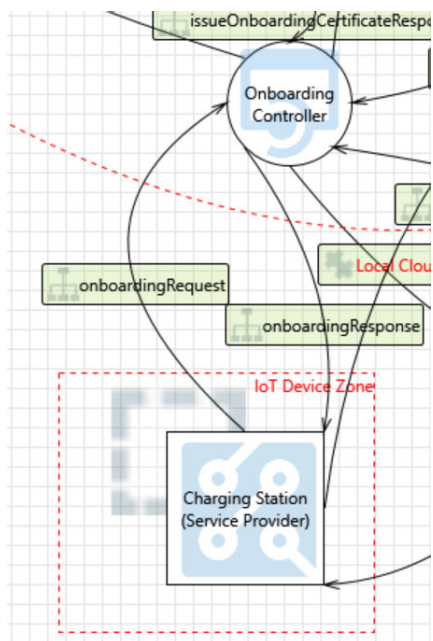

**FIGURE 24.** Data Flow Diagram. Interaction between the Charging Station (Service Provider) and Onboarding Controller system.

Since the identified threats violate primary security objectives, such as confidentiality and integrity, it is of utmost importance to investigate security standards and best practice guidelines for extracting security requirements that should be in place for mitigating them. Several security standards and best practice guidelines can be used to extract security requirements such as ISO 27000 series, NIST SP 800 series, ISA/IEC 62443 series.

We have selected ISA/IEC 62443 series, which are used as de facto standards for cybersecurity in the industrial IoT area. These standards are extensive and provide guidelines and requirements for the entire life-cycle such as production processes, systems, development processes, components and device security. For the purpose of this assessment, only one specific standard is considered, IEC 62443-3-3 *System security requirements and security levels*.[4] The standard provides and groups System Requirements (SRs) into seven Foundational Requirements (FRs): FR1 Identification and Authentication Control, FR2 Use Control, FR3 System Integrity, FR4 Data Confidentiality, FR5 Restricted Data flow, FR6 Timely Response to Events, and FR7 Resource Availability [51]. The identified threats are mapped with the corresponding SRs in IEC 62443-3-3 to identify possible mitigations. Then, the automated onboarding procedure is assessed against these SRs, to show how it can fulfil them for mitigating the threats.

The mapping of threats, SRs, and mitigations provided by the automated onboarding procedure is shown in Table 8. The SRs are mitigated in compliance with the IEC 62443-3-3 standard. Conformance with established security standards and best practice guidelines is essential for establishing a security baseline. For example, our approach provides a chain of trust from the hardware device, to its hosted application systems and their provided services by creating a chain of certificates to address [SR 1.2 RE 1],[SR 1.5],[SR 1.8], and [SR 1.9]. A new hardware device cannot be registered in DeviceRegistry without a valid *manufacturer certificate* and *Arrowhead onboarding certificate*, a system cannot be registered in SystemRegistry without a valid *device certificate*, and a service cannot be registered in ServiceRegistry without a valid *system certificate*. The device, system and service unique IDs and certificates are separately stored in the respective registries to increase security. Some specific threats, e.g. [T03] and [E02], are not applicable for the onboarding systems because they depend on the service provider. The same applies to the remaining 5 interactions shown in Figure 23, so the onboarding procedure can mitigate approximately 76 threats out of 88 threats in total. However, other support core systems of Eclipse Arrowhead can be integrated during the onboarding procedure to check the fulfillment of these requirements. For example, monitoring and standard compliance verification (MSCV) system can be used to check if the service provider prohibits the use of unnecessary functions, ports, protocols, and services [SR 7.7] [52]. The results of the security assessment show that the identified threats can be mitigated by the automated onboarding procedure based on the SRs defined in IEC 62443-3-3 standard.

## VII. CONCLUSION

In this paper we have discussed existing IoT frameworks and the security requirements that should be considered to protect against cyber threats, e.g. identity management, authentication and authorization, secure communication protocols. Since software-based security mechanisms are not sufficient to protect against existing security threats because data may be collected by potentially untrusted devices, we propose to add an additional hardware-based security layer via secure elements. We have proposed an automated onboarding procedure, which is used to create a chain of trust from the hardware device, to its hosted software systems and their services, whenever a new device wants to interact with an IoT/SoS framework. We implemented and evaluated it using the Eclipse Arrowhead framework and extended the initial concept presented in [7] to improve security and to address the complexity of devices and their needs. We have introduced the Onboarding Controller system, which is the first entry point to the Arrowhead local cloud and on success provides: (i) the Arrowhead issued ''onboarding'' certificate, and (ii) the endpoints of other services needed for the onboarding procedure. Additionally, we have introduced the CA system, which is responsible for signing any descendant certificates in an Arrowhead local cloud. All parties must trust the CA system registered with the common name of its hosting local cloud. We have shown the application of the onboarding procedure in an industrial use case and have performed a security assessment. The results have shown that the proposed procedure is compliant with the IEC 62443-3-3 security requirements and it improves performance, since the time required to run the use case is reduced compared with manual onboarding.

---
[4]https://webstore.iec.ch/publication/7033

**TABLE 8.** Security Assessment. Mapping of threats, IEC 62443-3-3 SRs, and mitigations provided by the automated onboarding procedure for the interaction between the charging station and Onboarding Controller system.

| Threat Category | Threats identified using Microsoft Threat Modeling Tool | IEC 62443-3-3 | Mitigation |
|---|---|---|---|
| Spoofing (S) | [S01] An adversary may spoof Charging Station (Service Provider) and gain access to Web API | [SR 1.2 RE 1] [SR 1.5] [SR 1.8] [SR 1.9] | Devices, systems and services get unique IDs and digital certificates (X.509) through the onboarding procedure, which are separately stored in DeviceRegistry, SystemRegistry and ServiceRegistry systems |
| Tampering (T) | [T01] An adversary may inject malicious inputs into an API and affect downstream processes | [SR 3.2 RE 1] | Model validation is done on Web APIs methods |
| | [T02] An adversary may tamper Charging Station (Service Provider) and extract cryptographic key material from it | [SR 1.5 RE 1] [SR 1.9 RE 1] | The automated onboarding procedure supports the integration of hardware security modules (e.g. TPM) |
| | [T03] An adversary may tamper the OS of a device and launch offline attacks | [SR 3.4] | MSCV system checks the OS and additional partitions of the device against unauthorized changes using cryptographic hashes |
| | [T04] An adversary can gain access to sensitive data by performing SQL injection through Web API | [SR 3.5] | All input information use type-safe parameters and is validated (enforced in code review) |
| Repudiation (R) | [R01] Attacker can deny a malicious act on an API leading to repudiation issues | [SR 2.12 RE 1] | Central auditing and TokenGeneration service of Authorization system are responsible for addressing this requirement |
| Information Disclosure (I) | [I01] An adversary can gain access to sensitive information from an API through error messages | [SR 3.7] | Exception handling is done in Web API and error messages do not disclose sensitive information |
| | [I02] An adversary can gain access to sensitive data by sniffing traffic to Web API | [SR 4.1] | The connection between Charging Station (Service Provider) and Onboarding Controller is secured using HTTPs |
| | [I03] An adversary can gain access to sensitive data stored in Web API's config files | [SR 4.1] | Partially mitigated, because only data in transit is encrypted |
| Elevation of Privileges (E) | [E01] An adversary may gain unauthorized access to Web API due to poor access control checks | [SR 2.1 RE 1] | Authentication is done through the manufacturer certificate. Once onboarded, Arrowhead LC deals with authorization using Authorization system |
| | [E02] An adversary may gain unauthorized access to privileged features on Charging Station (Service Provider) | [SR 2.1 RE 1] [SR 7.7] | Charging Station (Service Provider) as a precondition shall provide the capability to prohibit the use of unnecessary function, ports, protocols, and services. This requirement can be checked by the MSCV system |

As future work, we would like to implement a use case, in which an application system interacts with the generic autonomic management system (GAMS) [6] provided by Arrowhead, using the proposed onboarding procedure. The onboarding procedure allows to rely on the information on which "smart" decisions are being based. This would support self-adaptation, while ensuring a secure and trusted communication between the application system and the core systems of Eclipse Arrowhead local cloud.

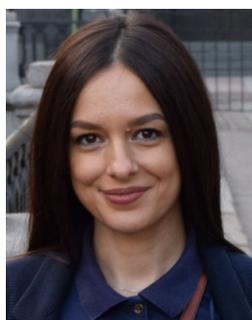

**SILIA MAKSUTI** received the Dipl.-Ing. degree in communication engineering from Carinthia University of Applied Sciences, Klagenfurt, Austria, and the B.Sc. degree in telecommunication engineering from the Polytechnic University of Tirana, Albania. She is currently pursuing the Ph.D. degree with Luleå University of Technology, Sweden. She also works as a Researcher with the Research Center for Cloud and Cyber Physical Systems Security, University of Applied Sciences Burgenland, Austria. Recently, she was working with Austrian Institute of Technology (AIT) in the AIT's ICT-Security Program. She has been part of several EU projects, such as SECCRIT, SEMI40, PRODUCTIVE4.0, ArrowheadTools, and Comp4Drones.

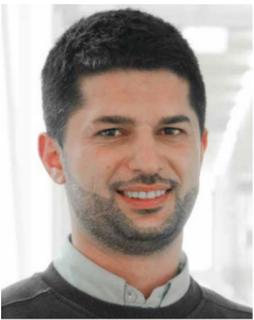

**ANI BICAKU** received the B.Sc. degree in telecommunication engineering from the Polytechnic University of Tirana, Albania, in 2012, the Dipl.-Ing. degree in communication engineering from Carinthia University of Applied Sciences, Austria, in 2015, and the Ph.D. degree in cyber-physical systems from Luleå University of Technology, in 2020. From 2014 to 2016, he worked with Austrian Institute of Technology (AIT), in the AIT's ICT-Security Program, and responsible for evaluating data security, data privacy, and high-assurance in cloud computing. In 2016, he joined the University of Applied Sciences Burgenland, Austria, as part of the Cloud and Cyber-Physical Systems Research Center. He is a member of Austrian Electrotechnical Committee (OEK) of the Austrian Electrotechnical Association (OVE) at IEC and CENELEC standardization bodies within TC65-WG10 "Security for Industrial Process Measurement and Control—Network and System Security." He has been part of several EU projects, such as SECCRIT, SEMI40, PRODUCTIVE4.0, ArrowheadTools, and Comp4Drones.

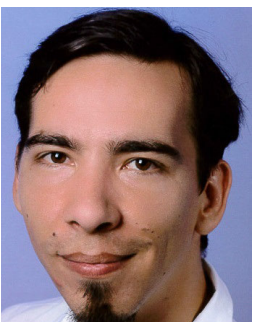

**MARIO ZSILAK** received the B.Sc. degree in information and communication systems and services from the University of Applied Sciences Technikum Wien, in 2017. He is currently pursuing M.Sc. degree in business process management and engineering with the University of Applied Sciences Burgenland. He also works as a Software Engineer with the Center for Cloud and CPS Security, Forschung Burgenland GmbH. He has contributed to arrowhead in a number of projects.

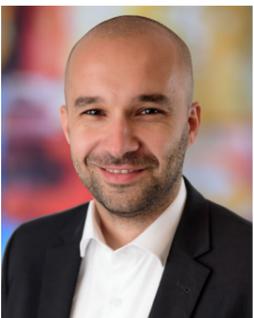

**IGOR IVKIC** received the B.Sc. degree in IT-infrastructure engineering and the M.Sc. degree in business process management engineering from the University of Applied Sciences Burgenland, Austria, in 2015 and 2017, respectively. He is currently pursuing the Ph.D. degree with Lancaster University, U.K. He also works as a Lecturer with the University of Applied Sciences Burgenland. He also works as a Researcher with the Research Center for Cloud and Cyber-Physical Systems Security (CCPSS), University of Applied Sciences Burgenland, in various projects, including MIT 4.0 (Project Leader) and Productive 4.0.

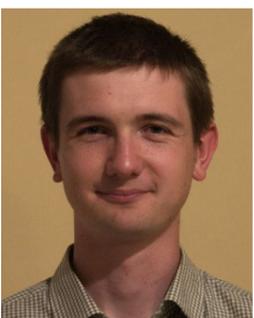

**BÁLINT PÉCELI** received the M.Sc. degree in electrical engineering from Budapest University of Technology and Economics (BME), Hungary, in 2016. Since 2015, he has been working with evopro Innovation Ltd., as a Research Software Engineer. He has been participated in European research programs, such as REPARA, ARROWHEAD, PRODUCTIVE4.0, and GREENERNET. His primary research interests include the IoT software engineering and the architectural design of modern industrial automation systems. He is also interested in the fields of high-performance computing and control engineering.

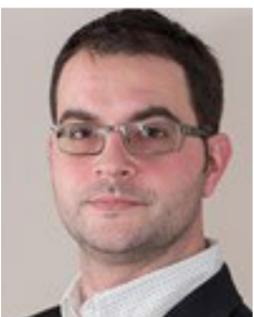

**GÁBOR SINGLER** received the B.Sc. degree in software system engineering from Széchenyi István University, Győr, Hungary, in 2005. Earlier, he was working mostly in process industry as a Process Engineer and a Software Engineer in SCADA system integration and software development. Later, he was working with Siemens A.G. MES Department, Genoa (system test), and Nuremberg (MES project consultant). After that, he worked with evopro Innovation Ltd., as a Research Software Engineer and a Software Theme Leader of evopro's AC and DC charging station development project, and an Operative Project Manager of a four year long reactor protection system testing software development project for the Hungarian Nuclear Power Plant. He is currently the Leader of the Industrial Software Development Business Unit, evopro Innovation Ltd. He participated in the ARROWHEAD European Research Program. His main research interests include electro mobility, energy management domain, the Internet of Things, and big physics.

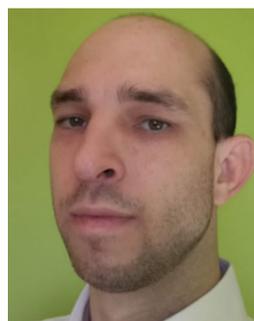

**KRISTÓF KOVÁCS** received the M.Sc. degree in electrical engineering from Budapest University of Technology and Economics (BME), Hungary, in 2016. Since 2015, he has been working with evopro Innovation Ltd., as a Research Software Engineer. He is currently working on Industrial Software Solutions as the Competence Team Leader. He participated in European research programs, such as REPARA, ARROWHEAD, and PRODUCTIVE4.0. He involved in embedded development of a gas spectrometer and electric vehicle chargers, communication and real-time module for a reactor protection test system for the Hungarian Nuclear Power Plant. His main interests include industrial embedded and distributed systems, communication networks, the IIoT, software architecting, and devops.

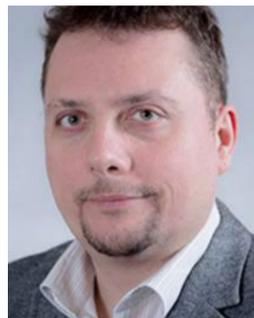

**MARKUS TAUBER** received the Ph.D. degree in computer science with a focus on autonomic management in distributed storage systems. He currently works as a Chief Scientific Officer with Research Studios Austria Forschungsgesellschaft mbH. From 2004 to 2012, he worked with the University of St Andrews, U.K., where he worked as a Researcher on various topics in the area of network and distributed systems. From 2012 to 2015, he coordinated the high assurance cloud research topic with Austrian Institute of Technology (AIT) part of AIT's ICT-Security Program. From 2015 to 2021, he worked as a FH-Professor with the University of Applied Sciences Burgenland, where he held the position as the director of the M.Sc. program in cloud computing engineering and led the Research Center Cloud and Cyber-Physical Systems Security. Amongst other activities, he was the coordinator of the FP7 Project Secure Cloud Computing for CRitical infrastructure IT—www.seccrit.eu, and involved in the ARTEMIS Project Arrowhead.

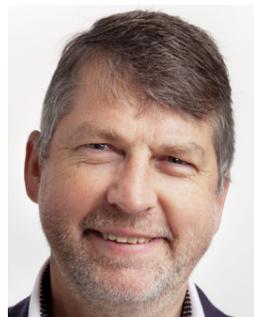

**JERKER DELSING** (Member, IEEE) received the M.Sc. degree in engineering physics from Lund Institute of Technology, Sweden, in 1982, and the Ph.D. degree in electrical measurement from Lund University, in 1988. From 1985 to 1988, he worked part time at Alfa-Lava—SattControl (now, ABB) with development of sensors and measurement technology. In 1994, he was promoted to an Associate Professor in heat and power engineering with Lund University. In early 1995, he was appointed as a Full Professor in industrial electronics with Lulea University of Technology, where he is currently the Scientific Head of EISLAB—http://www.ltu.se/eislab. He and his EISLAB Group has been a partner of several large EU projects in the field, such as Socrades, IMC-AESOP, Arrowhead, FAR-EDGE, Productive4.0, and Arrowhead Tools. His present research profile can be entitled the IoT and SoS automation, with applications to automation in large and complex industry and society systems. He is a Board Member of ARTEMIS, ProcessIT.EU, and ProcessIT Innovations.

● ● ●